\documentclass[prd,preprint,nofootinbib,showpacs,showkeys,amsmath,
amssymb,superscriptaddress,prd]{revtex4}%

\usepackage[dvips]{graphics}
\usepackage{array,amsmath,amssymb}
\usepackage{amsmath}
\usepackage{amssymb}

\newcommand{\dis}{\displaystyle}
\newcommand{\Tr}{\mbox{Tr}}

\begin{document}

\title{
The volume dependence of the long-range two-body potentials 
in various color channels by lattice QCD
}

\author{Y.~Nakagawa} 
\affiliation{Research Center for Nuclear Physics, 
 Osaka University, Ibaraki, Osaka 567-0047, Japan}

\author{A.~Nakamura}
\affiliation{Research Institute for Information Science and Education,
 Hiroshima University, Higashi-Hiroshima 739-8521, Japan}
 
\author{T.~Saito}
\affiliation{Integrated Information Center, Kochi University, Kochi, 780--8520, Japan}

\author{H.~Toki}
\affiliation{Research Center for Nuclear Physics, 
 Osaka University, Ibaraki, Osaka 567-0047, Japan}

\begin{abstract}
We study the color-dependent confining forces between two quarks by the
quenched lattice simulations of Coulomb gauge QCD.
The color-singlet and color-antitriplet instantaneous potentials yield
attractive forces. The ratio of the string tensions obtained from them
 is approximately 2 and have little volume dependence.
Meanwhile, the color-octet and color-sextet channels give a minor contribution
for two-quark system. We finally find that the infrared self-energy
of the color-nonsinglet channels diverges in the infinite volume limit;
 however, the degree of the divergence on the finite lattice can be understood
 in terms of color factors.
\end{abstract}



\maketitle

\section{Introduction}
The long-distance color-dependent forces among quarks and gluons
are the key quantities for the understanding of the internal
structure of hadron as well as color confinement
dynamics. The quarks and gluons are described 
in the quantum chromodynamics (QCD),
and have a color charge based on the SU(3) group,
with which the quark combination yields many color-dependent forces including
both attractive and repulsive forces. They make the
hadron structure (quark bound state) more
complicated. Therefore, it is essential
to know the basic behavior of the color-dependent
force at short and large distances.

The color-dependent force will be important 
when one investigates the multiquark hadron (made of more than four quarks)
and the exotic meson, etc. Although
many candidates of those particles have been recently
reported by experimental groups \cite{penta-e,Swanson}, 
it is still a hard task to theoretically understand those new particles.
They may consist of many quarks with various color-dependent
forces that make their internal structure more
complicated than the existing baryons.
The quark interaction at short distances can be
characterized by a coefficient $\left<\lambda\cdot\lambda\right>$ 
of a one-gluon exchange potential.
This is a basic assumption if we construct a quark model 
to deal with the baryon system. However, owing to the
quark confinement, it is not obvious how quarks behave
for large quark separation. We
thus need a lattice simulation to obtain a nonperturbative feature
 for the long-range color-dependent force.

However, the lattice study along this line is sparse,
although there are many lattice studies
about the $q\bar{q}$ potential \cite{Bali-r} obtained by
the gauge-invariant Wilson-loop operator.
This operator mixes color-singlet with color-octet contributions;
therefore, we cannot separately
extract the color-octet potential from the Wilson loop.
In spite of this, the color-octet
process has been discussed in the detailed analysis
of the $J/\psi$ photoproduction to explain
the deviation between the experimental data and
the theoretical predictions  \cite{Octet1,Octet2,Octet3}.
On the other hand, there is no lattice calculation
for $qq$ sector as well; however, the diquark (color-$3^*$ $qq$)
is also an important ingredient on hadron physics. 
The existence of the diquark has been believed by many physicists and applied
to work out unsolved problems \cite{Anselmino,Jaffe,
Nagata1,Nagata2,AbuRaddad,Nagata3}.
Note that the $qq$ potential in the lattice theory
is not formulated gauge-invariantly.

In this study, we employ a color-dependent Polyakov line correlator
(PLC) with the
Coulomb gauge fixing in order to clarify the color-dependent forces
\cite{Nakamura,Nadkarni}.
In the quark-gluon plasma phase,
there have been already some numerical studies of the color-screened
color-dependent force with the Coulomb and Landau gauge fixings
\cite{CDP1,CDP2,Maezawa,Kacz}.
Investigating
the confining color-dependent forces
we use Coulomb gauge QCD \cite{CCG},  
theoretical background of
which has been recently studied well \cite{CCG,PRC,Nie}.
In this theory, the PLC potential  
can be separated into the vacuum polarization part
 and the color-Coulomb instantaneous part;
 the latter relatively produces us a clear numerical signal
even in the confinement phase \cite{Greensite,Greensite2,Saito1,Saito2}.
The color-Coulomb instantaneous potential
 is not
the same as the Wilson-loop potential.
However, it yields an upper bound of a linearly-rising
quark potential \cite{Zwan} and
is the most important quantity on the Coulomb gauge confinement
scenario \cite{CCG}.
The infrared singularity of this potential 
is caused from accumulation of
the Faddeev-Popov (FP) ghost eigenvalue at the vanishing momentum
\cite{FP1,Nakagawa}.

Coulomb gauge QCD may provide us the most favorable framework
when one considers
the hadron phenomenological studies such as a constituent quark model.
The Coulomb gauge as a
physical gauge gives a positive-definite Fock space
that is suitable for constructing effective
Hamiltonians. Moreover, the color-Coulomb instantaneous potential
without the vacuum
polarization (or a retarded effect) is required to make 
a quark bound state in analogy with the quantum electrodynamics (QED).
 This formulation is applicable in the heavy-quark system
with no retarded effect as well as in the light-quark system 
with the constituent quarks due
to chiral-symmetry breaking \cite{Szc}. 

In this paper, we study the long-distance divergence behavior 
of the color-dependent forces. Because 
the color-nonsinglet state cannot exist in nature,
singularities in such states will emerge on the lattice 
as a finite-volume effect. 
This has not been verified in the previous
calculation [13], and thus we should investigate
their divergence behavior on a variety of
lattice sizes. We also discuss the color-factor dependence
for the degree of the
divergence and will confirm this point by the numerical
lattice simulation. In section II,
we briefly summarize Coulomb gauge QCD and
the definition of the color-dependent
potentials with the PLCs. In section III,
we give lattice numerical results and fitting analyses.
Section IV is devoted to our summary.

\section{Color-dependent potentials}

\subsection{Instantaneous potential in Coulomb gauge QCD}

Coulomb gauge QCD has been quantized through the Faddeev-Popov technique \cite{CCG}
and renormalizability of this theory 
has been also proved in terms of the Hamiltonian and Lagrangian formalism \cite{PRC,Nie}. 
The use of the Coulomb gauge as a physical gauge leads us to classify 
 transverse gluon modes and an instantaneous interaction,
 which is required to make quark bound states in analogy with QED.

The Hamiltonian of QCD in the Coulomb gauge can be given by 
\begin{equation}
\begin{array}{ccc}
H = \displaystyle
\frac{1}{2} \int d^3 x ( E_i^{tr2}(\vec{x})+B_i^2(\vec{x}) ) +
\displaystyle
\frac{1}{2} \int d^3 x d^3 y
(\rho(\vec{x}){\cal V}(\vec{x},\vec{y}) \rho(\vec{y}) ),
\label{part}
\end{array}
\end{equation}
where $E_i^{tr}$, $B_i$ and $\rho$ are
the transverse electric field, the transverse magnetic field and
the color-charge density, respectively.
The function $\cal V$ in the second term
is made by the Faddeev-Popov (FP) operator in the spatial direction,
$M=-\vec{D} \vec{\partial}=
-(\vec{\partial}^2 + g\vec{A} \times \vec{\partial})$,
\begin{equation}
{\cal V}(\vec{x},\vec{y})= \int d^3 z 
\left[ \frac{1}{M(\vec{x},\vec{z})}
( -\vec{\partial}^2_{(\vec{z})} )
\frac{1}{M(\vec{z},\vec{y})} \right].
\label{instinteraction}
\end{equation}

From the partition function with the Hamiltonian Eq. (\ref{part}),
 one can evaluate the time-time gluon propagator composed
 of the following two parts:
\begin{equation}
g^2\langle A_0(x) A_0(y) \rangle = 
g^2 D_{00}(x-y) = V (x-y) + P (x-y), 
\label{d00}
\end{equation}
where
\begin{equation}
V(x-y) = g^2 \langle {\cal V}(\vec{x},\vec{y}) \rangle \delta(x_4-y_4). 
\label{Vi}
\end{equation}
Equation (\ref{Vi}) is the instantaneous color-Coulomb potential
at equal time and causes antiscreening, so that this potential should be a confining potential
to attract quarks in hadrons 
and is the most important quantity on the Coulomb gauge confinement
scenario.
Note that Eq. (\ref{Vi}) in the case of
quantum electrodynamics (QED) as a nonconfining theory
is identified as a Coulomb propagator $\langle -1/\partial_i^2 \rangle$
or a Coulomb potential $ 1/r $.

Simultaneously,
Eq. (\ref{d00}) stands for a vacuum polarization,
\begin{equation}
P(x-y) = -g^2 
\langle 
\int {\cal V}(\vec{x},\vec{z})\rho(\vec{z},x_4) d^3 z 
\int {\cal V}(\vec{y},\vec{z}{\,'})\rho(\vec{z}{\,'},y_4) d^3 z' 
 \rangle,
 \end{equation}
 which brings about color-screening effect owing to the negative sign, 
 producing the reduction of a color-confining force.
Note that this term is associated with a quark-pair creation from vacuum when dynamical quarks exist.
Consequently, we can consider that 
the instantaneous interaction 
is somewhat more classical than the time-dependent vacuum part.

\subsection{Color-Coulomb instantaneous potential} 
One can define the color-dependent potentials on a lattice 
with the Polyakov line (PL) correlators\cite{Nadkarni}. In this study
 we separate the original potential obtained 
from the PL correlators 
into the color-Coulomb instantaneous and color vacuum polarization 
(retarded) parts with the Coulomb gauge fixing.
 Moreover, the color-Coulomb potential defined by the link-link 
 correlator as will be described below
 gives clear signals into practical numerical calculation
 even in the quenched lattice simulations. 

We employ a partial-length Polyakov line (PPL) which
 can be defined as \cite{Greensite,Greensite2}
\begin{equation}
L(\vec{x},T) = \displaystyle\prod_{t=1}^{T} U_0(\vec{x},t),
\quad T=1, 2, \cdots, N_t.
\end{equation}
Here $U_0(\vec{x},t)= \exp(iagA_0(\vec{x},t))$
is an $SU(3)$ link variable in the temporal direction and
$a$, $g$, $A_0(\vec{x},t)$ and $N_t$ represent the lattice cutoff,
the gauge coupling, the time component of a gauge potential and
the temporal-lattice size.
A PPL correlator in the color-singlet channel is given by
\begin{equation}
G_1(R,T) = \frac{1}{3}\left< \Tr[L(R,T)L^{\dagger}(0,T)] \right>,\label{pots}
\end{equation}
where $R$ stands for $\arrowvert \vec{x} \arrowvert $.
From Eq. (\ref{pots}) one evaluates a color-singlet potential on a lattice,
\begin{equation}
V(R,T) = \log \left[
\frac{G_1(R,T)}{G_1(R,T+a)} \right]\label{pot1}.
\end{equation}
For the smallest temporal-lattice extension, i.e., $T=0$,
we define
\begin{equation}
V(R,0) = - \log [ G_1(R,1) ]\label{pot2}.
\end{equation}
Here $V(R,0)$ in the Coulomb gauge is assumed to be 
 the color-Coulomb instantaneous potential $V_{coul}(R)$.
The $V(R,T)$ in the limit $T \rightarrow \infty$ 
becomes the usual Polyakov line correlator.
These two potentials are expected to satisfy
Zwanziger's inequality, $V_{phys}(R) \le V_{coul}(R)$ \cite{Zwan}, 
where $V_{phys}(R)$ is the physical potential extracted from 
the Wilson loop.

\subsection{Color-dependent potentials on a lattice}

We apply the above discussion to the other $SU(3)$ color-dependent potentials
 between two quarks \cite{Nadkarni}. 
A color-octet correlator on $q\bar{q}$ is given by
\begin{equation}
\begin{array}{ccl}
G_8(R,T)&=&\dis\frac{1}{8} \left< \Tr L(R,T)\Tr L^{\dagger}(0,T) \right>
        -\dis\frac{1}{24}\left< \Tr L(R,T)   L^{\dagger}(0,T) \right>,
         \label{octet}
\end{array}
\end{equation}
and $qq$ correlators in the symmetric-sextet and antisymmetric-triplet channels 
($3 \otimes 3 = 6 \oplus \bar{3}$) are also given as
\begin{equation}
\begin{array}{ccl}
 G_{6}(R,T)& = &\dis \frac{3}{4}\left<\Tr L(R,T)\Tr L(0,T)\right>
            + \dis \frac{3}{4}\left<\Tr L(R,T)   L(0,T)\right>,
          \label{symmetric}
\end{array}
\end{equation}
\begin{equation}
\begin{array}{ccl}
G_{\bar{3}}(R,T) &=&\dis \frac{3}{2}\left<\Tr L(R,T)\Tr L(0,T)\right>
                 -\dis \frac{3}{2}\left<\Tr L(R,T)   L(0,T)\right>.
         \label{antisymmetric}
\end{array}
\end{equation}
In the same way as described in Eqs. (\ref{pot1}) and (\ref{pot2}) 
we obtain the color-dependent potentials
in each color channel.

The above four potentials are classified in terms of
the color (the quadratic Casimir) factor on color $SU(3)$ group
 in the fundamental representation:
\begin{equation}
C_{q\bar{q}}^{1}=-\frac{4}{3},\hspace{0.5cm}
C_{q\bar{q}}^{8}= \frac{1}{6},\hspace{0.5cm}
C_{q     q }^{\bar{3}}= -\frac{2}{3},\hspace{0.5cm}
C_{q     q }^{6}= \frac{1}{3},\label{colorfac}
\end{equation}
for color-singlet, color-octet, color-triplet and color-sextet channels, respectively.
These coefficients appear as the proportional constant of 
the one-gluon exchange potential 
\footnote{
Here we intend to discuss an irreducible representation for $q\bar{q}$ and 
$qq$ sectors in relation to a so-called $\left<\lambda\cdot\lambda\right>$
quark model. Therefore, in this study, we do not calculate
the gauge-invariant Wilson-loop potentials in
higher representations for $q\bar{q}$ sector as introduced 
in Refs. \cite{Deldar,Bali2,Faber,Koma} }.

\subsection{Gauge fixing}

Since the color-dependent potentials defined by the PPL correlators
are not gauge invariant,
we have to fix a gauge. We use the Coulomb gauge
 realized on a lattice as
\begin{equation}
\mbox{Max} \sum_{\vec{x}} \sum_{i=1}^3
\mbox{ReTr} U_i^{\dagger}(\vec{x},t),
\end{equation}
by repeating the following gauge rotations:
\begin{equation}
U_i(\vec{x},t) \rightarrow U_i^{\omega}(\vec{x},t)
= \omega^{\dagger}(\vec{x},t)U_i(\vec{x},t)
\omega(\vec{x}+\hat{i},t),
\end{equation}
where $\omega$ $\in SU(3)$ is a gauge rotation matrix
\footnote{Here we used $\omega=e^{i\alpha \partial_i A_i}$,
where the parameter $\alpha$ is chosen suitably depending on the lattice size, etc.}
{} and
$U_i(\vec{x},t)$ are spatial lattice link variables.
The thermalized lattice configuration can be gauge fixed iteratively
\cite{Mandula}.

Because the Coulomb gauge fixing does not fully fix a gauge 
one can still perform a time-dependent gauge rotation
 on the Coulomb-gauge fixed links,
\begin{equation}
\begin{array}{ccl}
U_i(\vec{x},t) &\rightarrow&
\omega^{\dagger}(t) U_i(\vec{x},t) \omega(t), \\
U_0(\vec{x},t) &\rightarrow&
\omega^{\dagger}(t) U_0(\vec{x},t) \omega(t+1) \label{gt}.
\end{array}
\end{equation}
Thereby, $\mbox{Tr}L\mbox{Tr}L^{\dagger}$ and $\mbox{Tr}LL$
constructed by PPL
are not invariant under this transformation Eq. (\ref{gt}).
Accordingly, when performing numerical simulations
for the octet and two $qq$ correlators with 
$\mbox{Tr}L\mbox{Tr}L^{\dagger}$ and $\mbox{Tr}LL$,
we should additionally implement a global temporal-gauge fixing
on the Coulomb-gauge fixed links as
\begin{equation}
\mbox{Max } \frac{1}{V}\sum_{\vec{x},t} \mbox{ReTr} U_0^{\dagger}(\vec{x},t)
\mbox{ under Eq. (\ref{gt})}, \label{gtg}
\end{equation}
where $V=N_x N_y N_z$ is a spatial lattice volume.
Note that this gauge fixing
does not affect an intrinsic Coulomb gauge feature.

\subsection{Infrared divergence of the color-dependent potential}

In Coulomb gauge QCD, 
 we find that infrared divergences cancel only for color-singlet interactions.
Let us write the color-charge density for two-quark system as
$\rho_a \sim T_1^a \delta(x-x_0) + T_2^a\delta(x-y_0)$,
where $T_i^a$ are the generators of the color-$SU(3)$ group.
The instantaneous potential with $R=x_0-y_0$ (Eq. (\ref{instinteraction}))
 is given by 
\begin{equation}
V(\vec{R}) = T_1^aT_2^b
 \int \frac{d\vec{p}}{(2\pi)^3}
\frac{d^2(p)f(p)}{p^2} e^{i\vec{p}\cdot \vec{R}},   
\label{self1}
\end{equation}
where $d(p)$ is the expectation value of the Faddeev-Popov operator 
and $f(p)$ means the deviations of the factorization, e.g., $f(p)=1$ 
if there is no quantum correction; this argument has been done  
in Refs. \cite{FactorZwan,FactorSzczepaniak}. 
Here $T_1^a T_2^b$ is reduced to each color factor
 that appeared in Eq. (\ref{colorfac}).
On the other hand, the self-energy interaction 
not depending on the distance $R$ is written as 
\begin{equation}
\displaystyle
\Sigma = (T_i^a)^2
 \int \frac{d\vec{p}}{(2\pi)^3}
\frac{d^2(p)f(p)}{p^2}, 
\label{self2}
\end{equation}
where the Casimir invariant $(T_i^a)^2=4/3$
in the fundamental representation of the $SU(3)$ group.

The infrared (not ultraviolet) divergence emerges in 
both Eps. (\ref{self1}) and (\ref{self2}).
If the term $d^2(p)f(p)$ behaves as $(1/\sqrt{p})^2\cdot 1/p$ that 
would be responsible for the linear confinement $V \sim 1/p^4$, 
as has been analyzed in the Refs. \cite{FactorSzczepaniak,Langfeld},
then the infrared singularities unrelated to the linear potential
arise from two terms: 
\begin{equation}
V^{IS}(\vec{R}) = \displaystyle 4\pi (T_1^aT_2^b)  
\int_{0}^{\infty} dp \frac{1}{p^2},
\hspace{0.5cm}
\displaystyle
\Sigma^{IS} = \displaystyle 4\pi (T_i^a)^2 \int_{0}^{\infty} dp \frac{1}{p^2}.
\end{equation}
Consequently,
 they are completely cancelled in the case of the color-singlet representation 
since $(T_1^aT_2^b) + (T_i^a)^2 = (-4/3) + 4/3 = 0$.
Meanwhile, the other cases are proportional to the following factors: 
 $3/2$, $2/3$, $5/3$ for $8$, $3^*$ and $6$, respectively,
implying that the color-sextet channel may diverge most strongly.

\section{Numerical Results}

\subsection{Simulation parameters and statistics}

We carry out $SU(3)$ lattice gauge simulations
in the quenched approximation to calculate the color-decomposed PPL correlators.
The lattice update is done by the heat-bath Monte Carlo algorithm with a plaquette Wilson gauge action.
The lattice configuration numbers for the $18^4$, $24^4$ and $32^4$
 lattices are 600, 700 and 320; additionally, in order to investigate 
the volume dependence, we added the $8^4$ and $12^4$ lattices with 
$200$ configurations.
The lattice coupling constant $\beta$ for all the lattices is fixed to $5.9$
corresponding to the lattice cutoff
 $a \sim 0.12 fm$ \cite{qcdtaro}.

\subsection{Color-dependent potentials}

\begin{figure}[htbp]
\begin{center}
\resizebox{10cm}{!}{\includegraphics{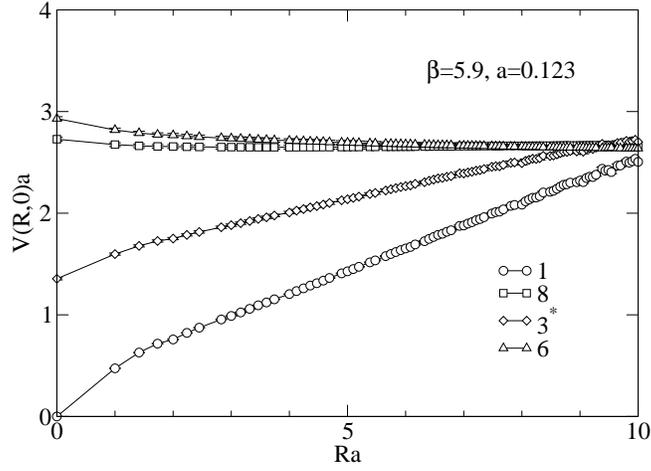}}
\caption{
Lattice numerical results of 
color-Coulomb instantaneous potentials between two quarks.
($\beta=5.9,a\sim 0.12fm$)}
\label{4pots}
\end{center}
\end{figure}

Figure \ref{4pots} shows 
 numerical results for the color-Coulomb instantaneous potential $V(R,0)$, 
 in the color-singlet, color-octet,  color-sextet, 
 color-triplet (antisymmetric) channels.
We find that both the color-singlet $V_1$ 
and color-antitriplet $V_{3^*}$ yield attractions at all distances,
and in particular, are linearly rising potential at large distances.
On the other hand, the color-octet $V_{8}$ and color-sextet $V_{6}$ potentials 
are repulsive forces although the variation of those potentials 
on distances is small.

\begin{figure}[htbp]
\begin{center}
\resizebox{10cm}{!}{\includegraphics{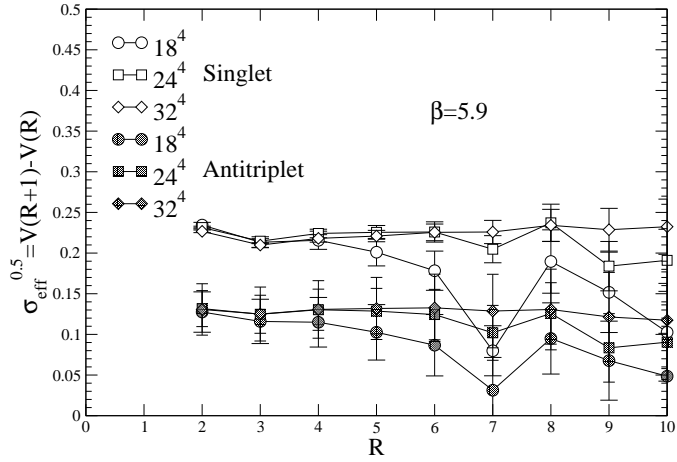}}
\caption{
Effective string tensions for the color-singlet and color-antitriplet channels 
in lattice units. 
Open (Filled) symbols are the color-singlet (antitriplet) string tensions.
Here the error bars are estimated in the error propagation.
The effective string tensions at large $R$ 
are strongly affected by the periodic boundary condition of a finite lattice; 
in contrast, 
owing to the increase of the lattice size, they become more stable even at large $R$. 
\label{effstring}
}
\end{center}
\end{figure}

\begin{figure}[htbp]
\begin{center}
\resizebox{10cm}{!}{\includegraphics{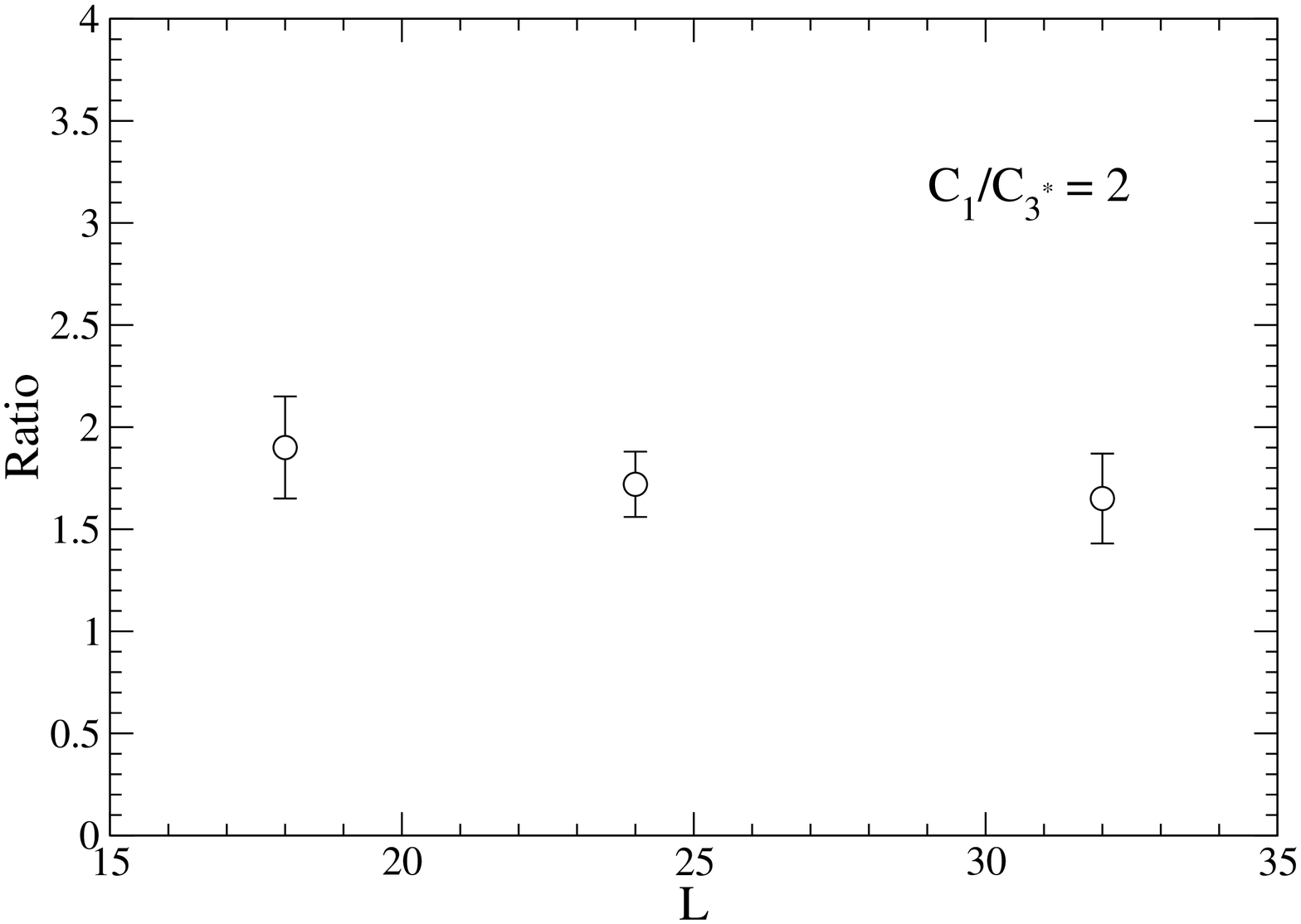}}
\caption{
Vertical axis stands for 
ratios $K_1/K_{3^*}$ of the string tensions
in the singlet and antitriplet channels and 
horizontal axis is the spatial lattice size.
\label{ratios}
}
\end{center}
\end{figure}

In order to investigate the magnitude of the string tensions, 
we calculate an effective string tension for the color-singlet and 
color-antitriplet channels.
The confining potential can be described by the function $V(R)=C+KR+A/R$ 
where $C$ is a self-energy constant term, K corresponds to the string tension, and 
the last term is the Coulomb term. 
Here the effective string tension is defined as 
$K=V(R+1)-V(R)$ in lattice units, which 
 should be a constant for large quark separations 
if it is a confining potential with a finite string tension.
In Fig. \ref{effstring} we find that the Ks become stable over approximately $R=3\sim4$ as the 
lattice size increases.
When we use the data for $R=3-6$ ($R=3-5$ for the $18^4$ lattice),
we plotted in Fig. \ref{ratios} the ratio $K_1/K_{3^*}$,
which is found to be close to $C_1/C_{3^*}=2$; 
for example, we obtain the $K_1=0.218(2)$ and $K_{3^*}=0.127(13)$ from 
the $24^4$ lattice.
Note that our definition of the color-Coulomb instantaneous part on a lattice 
 in terms of the PPL correlator (Eq. (\ref{pot2}))
does not completely exclude a vacuum polarization effect.
\footnote{A possible way to improve this discussion may be to construct directly
the instantaneous potential by the Faddeev-Popov propagator that has 
an infrared singularity\cite{Nakagawa,FP1}.
}
{}

For the color-octet and sextet channels it is hard to discuss
a precise value of both string
tensions under the present statistical accuracy.
In practice, from $24^4$ lattice data and under
the fitting range $R = 3-6$, we obtain the octet string tension
$K_8 = 0.0014(222)$ and the
sextet string tension $K_6 = -0.018(19)$
although the volume dependence does not appear
large as shown in Fig. 4.
This indicates that the long-range behavior of the octet and sextet
channels is quantitatively a minor contribution to two-quark system.
However, in order to work out more quantitative measurement
for the repulsive channels (and also the attractive
channels), one may need enough statistics on a larger lattice.

\begin{figure}[htbp]
\begin{center}
\resizebox{10cm}{!}{\includegraphics{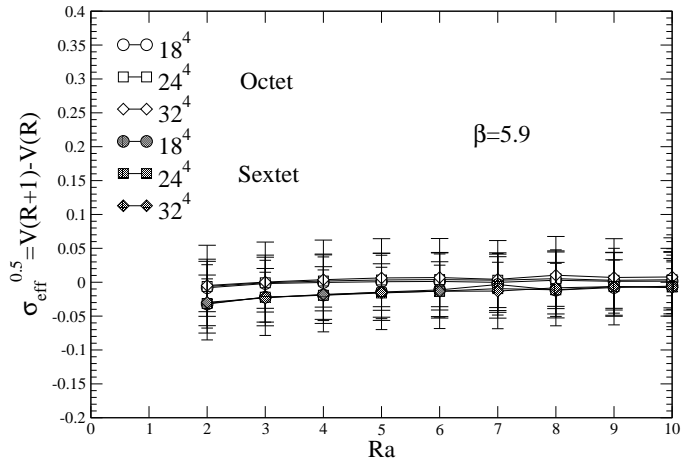}}
\caption{
Effective string tensions for the color-octet and color-sextet channels
in lattice units.}
\label{effstring2}
\end{center}
\end{figure}

\subsection{Divergence of color-nonsinglet potentials}

\begin{figure}[htbp]
\begin{center}
\resizebox{10cm}{!}{\includegraphics{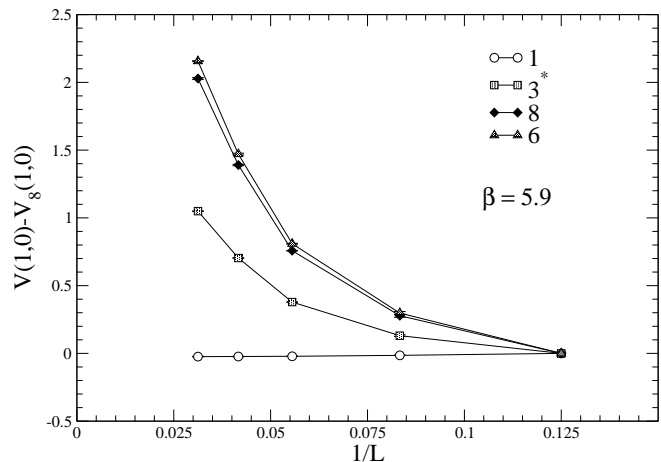}}
\caption{
Volume dependence of the color-dependent instantaneous 
potentials at the distance $R=1$ scaled by the $V^{8}(1,0)$ on the $8^4$ lattice.
The lattice sizes ($L$) used here are $8$, $12$, $18$, $24$ and $32$.
The variation of the singlet potential with the lattice volume 
is little while the other color-nonsinglet potentials diverge
 in the infinite volume limit.}
\label{Divfor4pot}
\end{center}
\end{figure}

\begin{figure}[htbp]
\begin{center}
\begin{minipage}{0.475\hsize}
\resizebox{7cm}{!}{\includegraphics{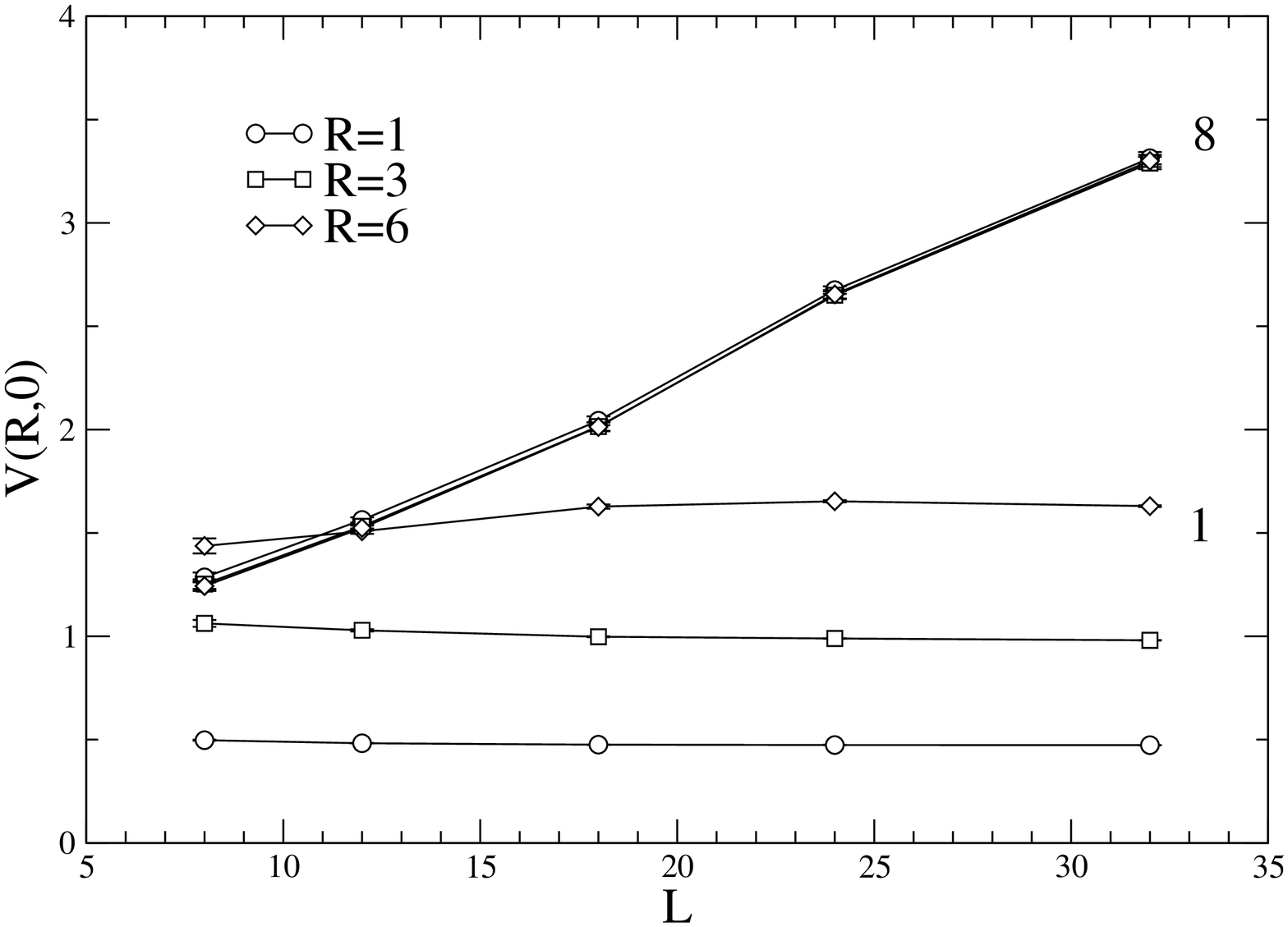}}
\end{minipage}
\hspace{-1.5cm}
\begin{minipage}{0.475\hsize}
\resizebox{7cm}{!}{\includegraphics{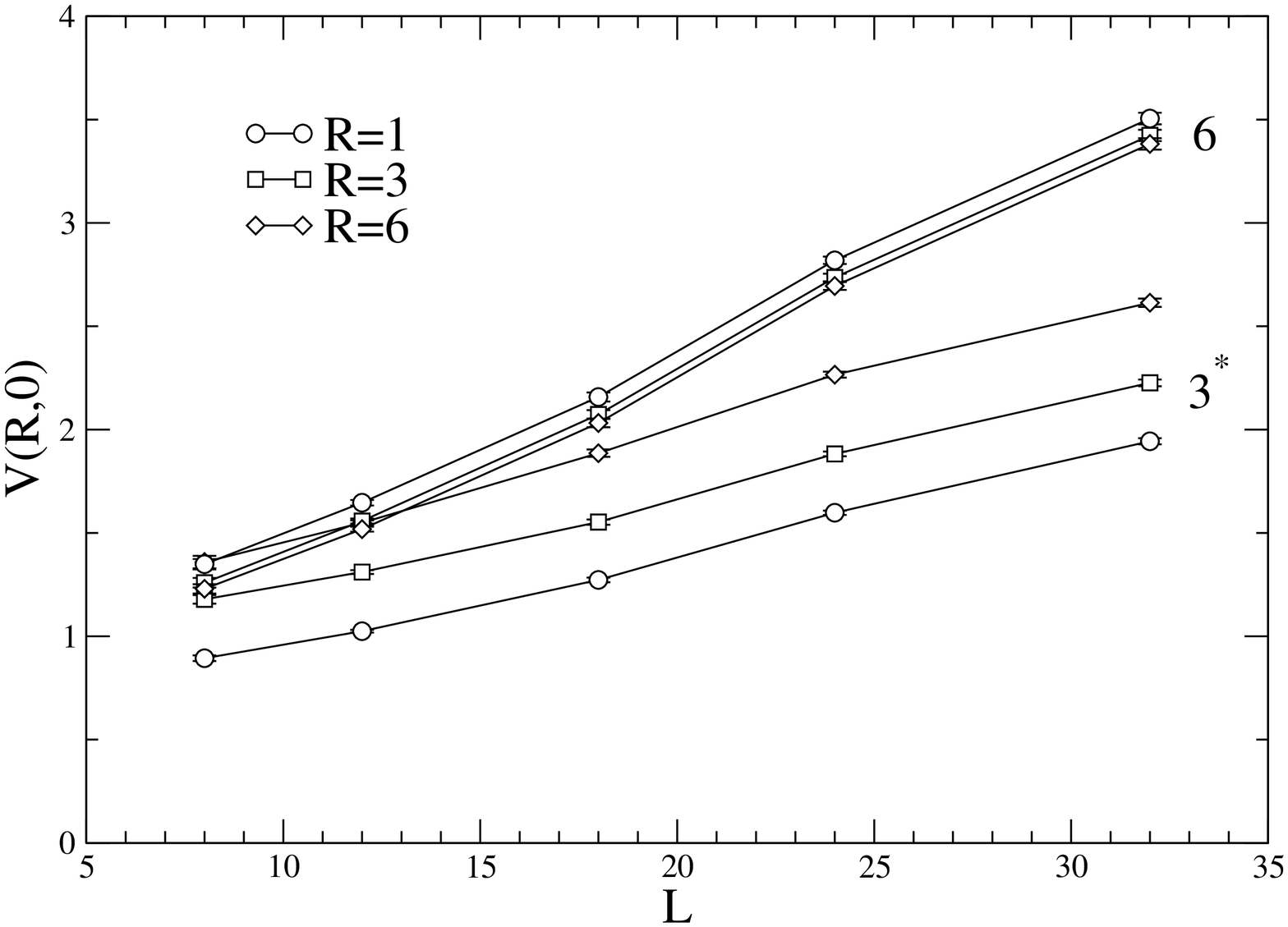}}
\end{minipage}
\caption{
Volume dependence of the color-dependent instantaneous 
potentials ($\beta=5.9$) at the distances $R=1,3,6$.}
\label{Divfor4pot2}
\end{center}
\end{figure}

Here we consider the infrared divergence property of
the color-dependent potentials while
the string tensions obtained from them
have little volume dependence as shown in the previous section.
The short-distance Coulomb term $\sim 1/R$,
which is not related to the color
confinement, may not matter on this argument.
Figure \ref{Divfor4pot} shows our nonperturbative numerical results,
by which we find that the color-singlet potential
has little volume dependence and the absolute value of
the color-nonsinglet potentials increases with the lattice volume;
this tendency remains at the distances $R = 3$ and $6$ as displayed
in Fig. \ref{Divfor4pot2}.
Thereby we conclude that the color-nonsinglet potentials diverge
regardless of the distance.

\begin{figure}[htbp]
\begin{center}
\resizebox{10cm}{!}{\includegraphics{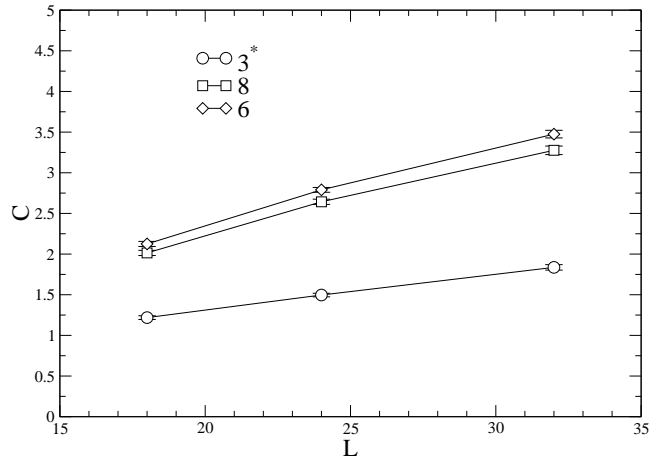}}
\caption{
Volume dependence of the self-energy term for color-nonsinglet potentials.
The self-energy
term diverges in the infinite volume limit.}
\label{DivC}
\end{center}
\end{figure}

Moreover, in order to estimate the degree of divergence
for the color-nonsinglet channels,
we assume the fitting function $f(R) = C + KR$ for
the potential data in the distance
$R = 3-6$ (long range). The corresponding self-energy term $C$s
fitted well are plotted in Fig. \ref{DivC};
this graph shows that the divergence contribution
increases monotonically with the volume.
Therefore we adopt the linear function $y = Dx+d$
where $D$ may correspond to the constant meaning the degree of divergence.
The fitted results are summarized in Table I,
which gives the conclusion that the magnitude of the divergence
on the finite lattice depends
on the ratio being $4:9:10$ for $3^*$, $8$ and $6$
as observed in the section II. E.
However, note that they will diverge equally to infinity
in the large volume limit,
as shown in Figs. \ref{Divfor4pot} and \ref{DivC}.
This indicates that the color-nonsinglet quarks cannot
exist independently; finally, they
will become a color-singlet state with finite energy to compensate
insufficient color degrees of freedom.

This numerical result does not contradict the dual-Ginzburg-Landau picture,
in which one understands that the color-singlet flux between two quarks
is shrunk like a string while the color-nonsinglet flux will radiate
rather than produce a closed string. A singularity
of the color flux not making the closed string can be expected
to emerge as the volume dependence on the lattice gauge theory.

\begin{table}[htbp]
\caption{Table of fitted results to estimate the degree of
divergence in color-nonsinglet channels.
$y = Dx + d$ is assumed as a fitting function
to fit the data of Fig. \ref{DivC}.}
\begin{center}
\begin{tabular}{c|c|c|c|}
    &$3^*$ & $8$ & $6$ \\
    \hline
$D$ &$0.044(06)$ & $0.092(14)$ & $0.098(14)$ \\
$d$ &$0.42(07)$  & $0.38(10)$  & $0.39(10) $ \\
$\chi^2/ndf$ & $0.19$ & $4.20$ & $4.63$ \\
\end{tabular}
\end{center}
\end{table}

\section{Summary}

We have studied 
the long-distance color-dependent forces
between two quarks in the quenched $SU(3)$ lattice simulation 
with the Polyakov line correlator.
We here focus the color-Coulomb instantaneous term 
in Coulomb gauge QCD, which has been discussed 
in the Gribov-Zwanziger confinement scenario and 
is required to make the hadron bound state consisting of quarks.

 Our numerical simulation shows that 
 the color-singlet $q\bar{q}$ channel as well as 
the color-antitriplet $qq$ (diquark) channel
causes a linearly confining potential at large distances.
The other color-octet and color-sextet channels at large distances yield
 also repulsive forces.
In addition, we find that the string tensions
 in the color-singlet and color-antitriplet
 channels have not significant volume dependence

We also investigated the infrared divergence of 
color-nonsinglet potentials and find that
the divergence on the finite lattice seems to be proportional
 to the color (Casimir) factor; however, the divergence of the
 color-singlet channel is not left over, In the infinite volume
 limit the color-dependent potentials except the color-singlet
 channel will diverge; in particular, the color-sextet channel
 diverges most strongly as expected. This conclusion is consistent
 with the dual-Ginzburg-landau picture of the color confinement.

This approach, in which we take notice of the color-Coulomb instantaneous part,  
may be suitable 
for further work to investigate three-quark color-dependent forces
for the understanding of baryons as well as new multiquark particles.
It is also necessary to study how the vacuum polarization term affects 
the instantaneous forces 
because it is reported in Refs. \cite{Nakamura,Philipsen1} that 
the octet (adjoint) channel calculated by
the full-length Polyakov line correlator
 with 
the vacuum polarization gives the complicated distance dependence 
at large distances. 

\section{Acknowledgments} 

The simulation was performed on SX-5 and SX-8 (NEC) vector-parallel computers 
at the RCNP of Osaka University. 
We appreciate the warm hospitality and support of the RCNP administrators.

\section{Appendix}

In Eq. (\ref{self1}), if the Coulomb kernel $K(p)$ produces a confining force, 
it is reported in Ref. \cite{Langfeld} that
the ghost factor $d(p)$ and the Laplacian factor $f(p)$
 behave as approximately $1/\sqrt{p}$ and $1/p$, respectively.
Thus we assume that $K(p)$ is $1/p^2$.
Omitting $T_1^aT_2^b$ term one also calculates Eq. (\ref{self1}):
\begin{equation}
\begin{array}{ccl}
V(\vec{R}) & = & \displaystyle 
 \int \frac{d\vec{p}}{(2\pi)^3} \frac{K(p)}{p^2} e^{i\vec{P}\vec{R}} \\
&=& \displaystyle
\int_0^{2\pi} \int_{-1}^{1}dcos\theta \int_0^{\infty} p^2 dp \theta\frac{K(p)}{p^2}
e^{ipRcos\theta} \\
&=& \displaystyle \frac{2\pi}{iR}
\int_0^{\infty} dp \frac{K(p)}{p} ( e^{ipR} - e^{-ipR} )\\
& = & \displaystyle \frac{2\pi}{iR}
\int_{-\infty}^{\infty} dp \frac{K(p)}{p} e^{ipR} \\
& = & \displaystyle \frac{2\pi}{iR}
\int_{-\infty}^{\infty} dp \frac{K(p)}{p}
( 1 + ipR + \frac{1}{2}(ipR)^2 + \frac{1}{3}(ipR)^3 + \cdots ).\\ 
\end{array}
\end{equation}
We here do the Taylor expansion in the infrared limit $p=0$ for $e^{ipR}$.
 In the above equation, the first term disappears
because it is a odd function on $p$; the second term proportional to $1/p^2$ 
causes the infrared divergence; the third term produces a (linear) potential; the other terms 
are irrelevant.
Finally, the second term relating to the divergence is  
\begin{equation} 
\begin{array}{ccl}
V^{IS}(\vec{R}) &=& \displaystyle 2\pi (T_1^aT_2^b)  \int_{-\infty}^{\infty} dp \frac{1}{p^2}\\
               &=& \displaystyle 4\pi (T_1^aT_2^b)  \int_{0}^{\infty} dp \frac{1}{p^2}.
\end{array}
\end{equation} 
In the same way, the relevant contribution 
 from the self-energy of Eq. (\ref{self2}) is also given by 
\begin{equation}
\Sigma^{IS} = \displaystyle 4\pi (T_i^a)^2 \int_{0}^{\infty} dp \frac{1}{p^2}.
\end{equation}

\end{document}